# LISA (LOCALHOST INFORMATION SERVICE AGENT)

**Iosif C. Legrand ***

**Ciprian Mihai Dobre\*\*, Ramiro Voicu\*\*, Corina Stratan\*\*,
Catalin Cirstoiu\*\*, Lucian Musat\*\***

*\* California Institute of Technology*
*\*\* "Politehnica" University of Bucharest*

Grid computing has gained an increasing importance in the last years, especially in the academic environments, offering the possibility to rapidly solve complex scientific problems. The monitoring of the Grid jobs has a vital importance for analyzing the system's performance, for providing the users an appropriate feed-back, and for obtaining historical data which may be used for performance prediction. Several monitoring systems have been developed, with different strategies to collect and store the information. We shall present here a solution based on MonALISA, a distributed service for monitoring, control and global optimization of complex systems, and LISA, a component application of MonALISA which can help in optimizing other applications by means of monitoring services. The advantages of this system are, among others, flexibility, dynamic configuration, high communication performance.

## 1. LISA FRAMEWORK

LISA (Localhost Information Service Agent) is one of the component applications of MonALISA, a larger project that provides a distributed monitoring system. This application helps in optimizing others applications by means of monitoring services. For that LISA was implemented in a centric manner in which a core component performs all the monitoring and on top of which numerous other applications can act as listeners for the obtained results.

A first advantage of LISA is that it can operate independent of the operating system platform. For that purpose Java was chosen for most of the parts. There are still parts written in C which are connected to the hole using the JNI technology. As a result LISA can run on Linux (with different versions of kernels, including 2.6), Windows and MacOS.

In the core LISA is composed of several monitoring parts which together help in best describing the local system current functioning parameters. Even better, these monitoring modules can be easily extended and others can be added as well.

The first monitoring part is the system module. This module offers environmental information such as the type of operating system and version, the user under which LISA is run, JVM version, local IP address (or even if the address is a private one the public address which stands for point of access to the local network), the AS to which the station belongs and so on. Another module is the host module. This is the most complex module because it offers information regarding the global running parameters of the local station. This information is classified into information regarding CPU utilization (how much of the CPU is used by the user applications, by the system or is idle), regarding memory utilization (free memory and total memory available in the system, page swapping), regarding disk utilization (free disk space and total disk space), regarding the current load of the system (load1, load5, load15) or the number of running processes and information regarding networking (in and out for each available network interface). LISA also contains a module that monitors the hardware configuration of the local station. And then another module which estimates the available bandwidth between the local station and some other end-point station from some other location. Also because LISA is a component part of MonALISA it contains a module which can interrogate repositories and classified farms in terms of connectivity to the local station network.

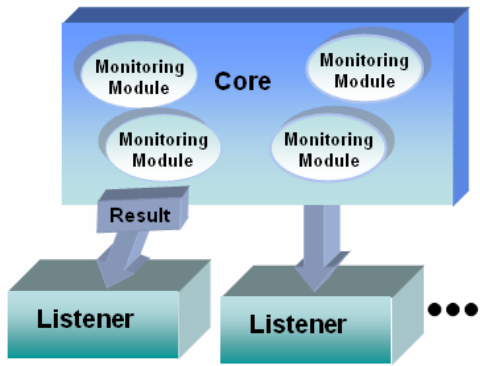

Fig. 1. LISA components and their interactions.

These modules are highly configurable and completely independent of each others. Because of this LISA does not require to run with all the modules performing monitoring at any time which means that every module can be started or stopped at any time. Also, as mentioned above, any new module can be easily added into the application. All this is meant to make LISA an application which best describes the local station in order to help other application in optimization from an end-user point of view.

The way in which LISA helps application is by sending the monitored parameters in results to a list of listener applications. Any application can implement the LISA listener interface and can register itself with LISA so to receive the monitored results. In addition LISA already has a number of modules which are listener to the core part and which can be viewed as example implementations. These are only examples, which mean that LISA can very well function without them, but can be useful in many ways.

The first listener module is the GUI client. The GUI can be used to visualize in a friendlier manner the current values of the monitored parameters for the local station. The GUI part can be viewed in the Java Web Start[1].

---

[1]LISA web page:http://monalisa.cacr.caltech.edu/lisa.
[2]ApMON web page:
http://monalisa.cacr.caltech.edu/apmon.

The GUI client itself is composed of two parts, one which shows the monitored results in charts and pies and another which shows the monitored results in a text based frame.

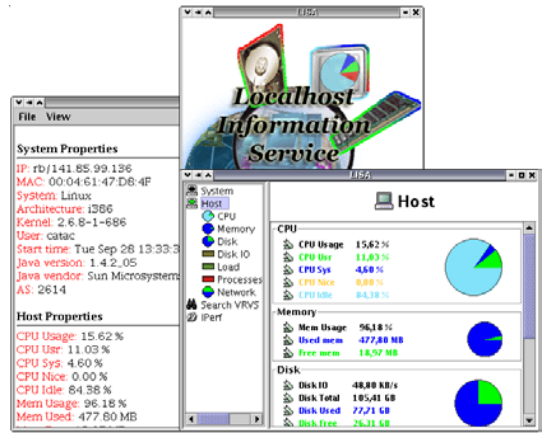

Fig. 2. The GUI client of LISA.

Because LISA is a part of MonALISA there is another module which sends the monitored values back to one or more MonALISA service. For that this modules uses the ApMON module[2]. This module ensures an even tighter bond between LISA and MonALISA. And because the latter is a much more complex system this module can help in that the monitored result values can be processed at a higher level using the MonALISA inner functions.

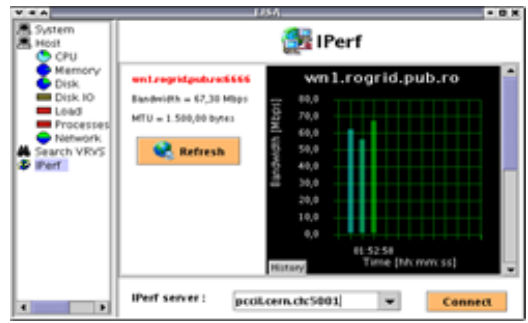

Fig. 3. Bandwidth measurement example using LISA.

As an example of the way LISA can help other applications we helped in assisting the VRVS client module to dynamically detect the best reflector to which to connect to. The module which helps in MonALISA peers discovery is the one which provided this functionality to the VRVS application. The best reflectors are read from MonALISA repositories and are updated in time so that the choosing is performed only from the available reflectors. Then the best reflectors are chosen based on their network location (Network domain, AS domain, Country, Continent) and on their current load values, number of currently connected clients, current network traffic. This insures a load balancing order in choosing the reflectors. Based on this from all the available reflectors the module that chooses some and performs RTT values measurements. In the end the module is able to inform some application (e.g. VRVS client) to reconnect to the most appropriate module if network conditions change and the quality is affected.

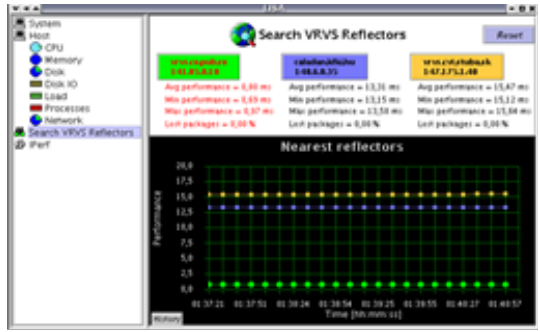

Fig. 4. LISA can help the VRVS client module to dynamically detect the best reflector to which to connect to.

In conclusion LISA, as part of the MonALISA project, is an application which can help by means of monitoring local station functioning parameters other end-user applications in many ways (optimization for instance). It can run itself as a standalone application which monitors the local station or can run as a sub-component of other end-user application.

REFERENCES

LISA web page:
    http://monalisa.cacr.caltech.edu/lisa.
ApMON web page:
    http://monalisa.cacr.caltech.edu/apmon.
MonALISA web page:
    http://monalisa.cacr.caltech.edu.
I.C.Legrand, "*MonALISA – MONitoring Agents using a Large Integrated Service Architecture*", Japan, 2003
H.B.Newman, I.C.Legrand, P.Galvez, R.Voicu, C.Carstoiu, "*MonALISA: A Distributed Monitoring Service Architecture*", CHEP 2003
Virtual Rooms VideoConferencing System:
    http://www.vrvs.org.
IPERF web page:
    http://dast.nlanr.net/Projects/Iperf/iperfdocs_1.7.0.html